\documentclass[lettersize,journal]{IEEEtran}

\usepackage{
  color,
  float,
  epsfig,
  wrapfig,
  graphics,
  graphicx
}
\usepackage{bm}
\usepackage{amssymb}
\usepackage{amsmath}

\usepackage{xparse} 
\usepackage{xspace}
\usepackage{makecell}
\usepackage{hyperref}
\usepackage{balance}
\usepackage{subfigure}
\usepackage{graphicx}
\usepackage{dblfloatfix}
\usepackage{epstopdf}
\usepackage{bbding}
\usepackage{pifont}
\usepackage[nointegrals]{wasysym}
\usepackage{rotating}
\usepackage{booktabs,caption}
\usepackage[flushleft]{threeparttable}
\usepackage{tabularx}
\usepackage{color}
\usepackage{xcolor}
\usepackage{colortbl}
\usepackage{soul}
\usepackage{comment}
\usepackage{url}
\usepackage{wrapfig}
\usepackage{tikz}
\usepackage{multirow}
\usepackage[ruled,vlined,linesnumbered]{algorithm2e}

\newenvironment{packeditemize}{
\begin{list}{$\bullet$}{
\setlength{\labelwidth}{8pt}
\setlength{\itemsep}{2pt}
\setlength{\leftmargin}{\labelwidth}
\addtolength{\leftmargin}{\labelsep}
\setlength{\parindent}{0pt}
\setlength{\listparindent}{\parindent}
\setlength{\parsep}{0pt}
\setlength{\topsep}{3pt}}}{\end{list}}

\begin{document}

\title{Security of World-Model-Based Embodied AI: A Lifecycle of Threats, Defenses, and Evaluation}

\author{Fazhong~Liu,
        Zhuoyan~Chen,
        Haozhen~Tan,
        Yan~Meng,
        Guoxing~Chen,
        and Haojin~Zhu,~\IEEEmembership{Fellow,~IEEE}
}

\maketitle

\begin{abstract}

World models give embodied AI a predictive core: they compress observations into states, simulate action-conditioned futures, and enable planning beyond reactive control. This predictive layer, however, opens a new security boundary-compromise can propagate from data, sensors, prompts, or feedback into physical action. Rather than treating world models as an isolated component, this survey traces threats across their entire lifecycle-from data construction and representation learning, through state grounding and imagination, to trajectory evaluation, execution, and long-term adaptation via memory and tools. We show that familiar attack families: poisoning, backdoors, adversarial examples, sensor spoofing, prompt injection, trajectory manipulation, and supply-chain attacks take on distinct meanings when they corrupt world states, learned dynamics, affordance estimates, or safety costs. We also highlight a duality: world models can serve as runtime safety shields, yet when compromised or over-trusted they generate predictive safety illusions. The survey offers a lifecycle taxonomy, maps existing attacks to world-model security properties, outlines evaluation protocols for safety failures, and structures defenses across provenance, robust grounding, uncertainty-aware prediction, trajectory gating, feedback auditing, and deployment assurance.

\end{abstract}


\newcommand{\etal}{\emph{et al.}\xspace}
\newcommand{\etc}{\emph{etc}\xspace}
\newcommand{\ie}{i.e.\xspace}
\newcommand{\eg}{e.g.\xspace}
\newcommand{\red}[1]{\textcolor{red}{#1}}
\newcommand{\yh}[1]{\textcolor{blue}{#1}}
\newcolumntype{Y}{>{\raggedright\arraybackslash}X}
\newcolumntype{P}[1]{>{\raggedright\arraybackslash}p{#1}}
\newcolumntype{C}[1]{>{\centering\arraybackslash}p{#1}}
\newcommand{\cmark}{\ding{51}}
\newcommand{\xmark}{\ding{55}}
\newcommand{\fSP}{\textsc{SP}}
\newcommand{\fSD}{\textsc{SD}}
\newcommand{\fPA}{\textsc{PA}}
\newcommand{\fNC}{\textsc{NC}}
\newcommand{\fPI}{\textsc{PI}}
\newcommand{\frnote}{\textsc{SP}: semantic-to-simulation-to-action gap; \textsc{SD}: state- and uncertainty-conditional risk; \textsc{PA}: rollout-to-execution amplification; \textsc{NC}: trajectory-level non-compositionality; \textsc{PI}: predictive safety illusion.}
\newcommand{\fullcov}{\CIRCLE}
\newcommand{\halfcov}{\LEFTcircle}
\newcommand{\partcov}{\Circle}
\newcommand{\dashcov}{--}

\section{Introduction}
\label{sec:introduction}


Embodied artificial intelligence is moving from reactive perception and scripted control toward agents that can predict, simulate, and evaluate possible futures before acting. This transition is visible in model-based reinforcement learning and latent imagination systems such as World Models and Dreamer \cite{ha2018worldmodels,hafner2020dreamer,hafner2023dreamerv3}, in world-model-based safe reinforcement learning \cite{safedreamer2023,vlsafe2025}, in video and interactive generative environments \cite{genie2024}, and in foundation-model-driven robots such as PaLM-E, RT-1, RT-2, OpenVLA, and Octo \cite{driess2023palme,brohan2022rt1,zitkovich2023rt2,kim2024openvla,octo2024}. In these systems, the agent no longer only maps the present observation to an immediate action. It maintains an internal representation of the world, imagines action-conditioned futures, and chooses behavior according to predicted feasibility, reward, cost, and safety.

This predictive capability creates a new security boundary. A compromised camera, LiDAR, prompt, map, simulator, video generator, memory store, policy adapter, or tool call can corrupt the world state or future rollout that the agent uses for decision making. In a conventional digital system, such corruption may cause incorrect classification or text generation. In embodied AI, it can cause unsafe physical motion, collision, rule violation, or harmful human-robot interaction \cite{wang2025exploring,sato2023lidar,zhang2025badrobot}. The security problem is therefore not only whether an input or model output is malicious, but whether a corrupted internal world model makes a dangerous future appear safe.


Existing surveys on embodied AI safety and security provide broad taxonomies of perception, language, planning, VLA, robotics, and cyber-physical threats \cite{liu2025aligning,xing2025towards,ma2026breaks,li2025safety,khan2024securi}. Other recent work has begun to discuss world-model-specific safety concerns \cite{baraldi2025safetychallenge,parmar2026worldrisks}. However, two gaps remain. First, world-model attacks are still scattered across literatures: adversarial perception, data poisoning, video generation, model-based RL, safe RL, VLA security, autonomous driving, robotics middleware, and agentic memory are often studied separately. Second, world models are usually treated either as an attack target or as a safety tool, but not as both. A world model may be attacked directly, used to amplify poisoned synthetic data, or deployed as a runtime safety checker whose own failure produces a false certificate of safety. This survey addresses both gaps by unifying these scattered literatures under a single lifecycle of world-model-mediated decision making, and by analyzing world models simultaneously as attack targets, poisoning amplifiers, and safety mechanisms.

We focus on \emph{world-model-based embodied AI}: systems that explicitly or implicitly use learned state representations, predictive dynamics, action-conditioned simulation, or long-term world knowledge to support embodied action. This scope includes explicit world-model agents, world-action models, safe model-based RL, video world models for driving or manipulation, and VLA/agentic robots whose policies rely on implicit physical and affordance models. The goal is not to replace capability-centric surveys. Instead, we reorganize established attack families around the lifecycle of world-model-mediated embodied decision making.


Five recurring insights structure the survey. We use the following abbreviations throughout the paper: \textsc{SP} for semantic-to-simulation-to-action gap, \textsc{SD} for state- and uncertainty-conditional risk, \textsc{PA} for rollout-to-execution amplification, \textsc{NC} for trajectory-level non-compositionality, and \textsc{PI} for predictive safety illusion.

\textbf{\fSP: Semantic-to-simulation-to-action gap.} A command, plan, or generated future can be semantically plausible yet physically unsafe. Textual refusal, symbolic plan checking, or visual quality of imagined video is insufficient unless the resulting trajectory is dynamically feasible and constraint-compliant \cite{jailbreaking_llm_controlled_robots_icra2025,li2024poex,physcondwma2026}.

\textbf{\fSD: State- and uncertainty-conditional risk.} Attack success depends on pose, viewpoint, lighting, object layout, map state, embodiment, and model uncertainty. A trigger or perturbation can remain benign in one state and become hazardous in another \cite{wang2025state,wang2024systematic}.

\textbf{\fPA: Rollout-to-execution amplification.} Small corruption in state estimation, latent dynamics, prompt interpretation, or action chunks can compound across imagined rollout and real closed-loop execution \cite{zhou2025annie,xu2025silentdrift,cao2024first}.

\textbf{\fNC: Trajectory-level non-compositionality.} Locally safe steps do not guarantee a safe long-horizon trajectory. World-model planning ranks multi-step futures, so attacks can target trajectory ranking rather than single-step prediction \cite{trap2026,zhou2025safebeam}.

\textbf{\fPI: Predictive safety illusion.} When a world model is used as a safety checker, an incorrect but confident prediction can make an unsafe action appear certified. This is especially dangerous for runtime monitors and generated future visualizations that humans or policies may over-trust \cite{safedreamer2023,huang2024guardrails}.


This survey makes four contributions.

\begin{packeditemize}
    \item We define a lifecycle framework for world-model-based embodied AI, covering data construction, training, state grounding, imagination, trajectory evaluation, execution feedback, and agentic extension.
    \item We map established attack families from adversarial ML, VLA security, video generation, robotics, CPS, and agentic AI into world-model security objects: state integrity, dynamics fidelity, affordance correctness, constraint compliance, trajectory-ranking integrity, uncertainty calibration, and feedback provenance.
    \item We analyze world models as both attack targets and safety mechanisms, highlighting generative world models as persistent poisoning sources and runtime safety world models as emerging attack surfaces.
    \item We synthesize evaluation and defense requirements for measuring and mitigating world-model-mediated physical risk, including unsafe rollout acceptance, predicted-safe-but-unsafe behavior, long-horizon cost, detectability, recovery latency, and sim-to-real degradation.
\end{packeditemize}

The remainder of this paper is organized as follows. Section~\ref{sec:background} introduces world-model-based embodied AI and the security objects used throughout the survey. Section~\ref{sec:taxonomy} presents the lifecycle framework and threat mapping. Sections~\ref{sec:preaction-threats} and \ref{sec:action-threats} analyze threats before and after action selection, including attacks on world models used as runtime safety checkers. Section~\ref{sec:evaluation} reviews benchmarks and metrics, Section~\ref{sec:defense} organizes defenses along the same lifecycle, and Section~\ref{sec:challenges} discusses open challenges and concludes the survey.

\section{Background}
\label{sec:background}

\subsection{Embodied AI and Predictive Cognition}

Embodied AI refers to agents that perceive, reason, and act in physical or simulated environments \cite{duan2022survey,argall2009survey}. Their decisions are coupled to sensors, controllers, actuators, humans, and surrounding infrastructure. This coupling makes security failures safety-critical: a corrupted observation, unsafe plan, or compromised controller can become physical harm rather than only digital error \cite{xu2024llm,liu2025aligning}.

World models give embodied agents a predictive layer between perception and action. At a minimum, a world model represents the current state and predicts how the environment changes under candidate actions. In model-based reinforcement learning, this role is explicit: latent dynamics models support imagination and planning \cite{ha2018worldmodels,hafner2020dreamer,hafner2023dreamerv3}. In foundation-model-driven robots, the world model may be implicit in a VLM, VLA, or LLM planner that has learned object semantics, affordances, and commonsense physical regularities from large-scale data \cite{driess2023palme,kim2024openvla,octo2024}. In agentic systems, long-term memory, maps, retrieved documents, tools, and execution logs become part of the agent's extended world knowledge \cite{wang2024voyager}.

\subsection{Forms of World-Model-Based Embodied AI}

    \textbf{Explicit latent world models.} Systems such as World Models, PlaNet, and Dreamer learn compact latent states and transition models, then plan or learn policies through imagined rollouts \cite{ha2018worldmodels,hafner2019planet,hafner2020dreamer}. DayDreamer later showed that the same latent-imagination recipe transfers to real robots \cite{wu2022daydreamer}. Safe model-based variants use predicted future cost or constraint violations to improve safety \cite{berkenkamp2017safe,safedreamer2023}.

\textbf{Generative world models and simulators.} Video or interactive generative models can synthesize future observations or simulated environments. Genie illustrates the idea of generating interactive environments from visual data \cite{genie2024}. Such models can support data generation and stress testing, but they also inherit attack surfaces from diffusion and video generation, including prompt-specific poisoning, backdoors, and adversarial temporal perturbations \cite{nightshade2023,badvideo2025,t2vattack2025}.

\textbf{World-action models.} Newer systems couple future prediction and action generation, making the imagined world and executable action tightly linked. This improves closed-loop capability but introduces alignment risks between what the model appears to predict and what it actually commands, as highlighted by emerging world-action drift attacks \cite{badwam2026}.

\textbf{VLA and implicit world models.} VLA policies such as RT-1, RT-2, OpenVLA, and Octo directly map language and observations to robot actions \cite{brohan2022rt1,zitkovich2023rt2,kim2024openvla,octo2024}. Even when no explicit rollout module is exposed, such models encode implicit beliefs about objects, actions, and task dynamics. VLA attacks and backdoors therefore become relevant to world-model-based embodied AI whenever they corrupt state grounding, action selection, or feedback updates \cite{liu2025attackvla,zhou2025badvla,li2025tabvla}.

\textbf{Hybrid stacks.} Practical systems often combine a VLM for perception, an LLM for task decomposition, a world model or simulator for prediction, a VLA or controller for action, and middleware for integration. SayCan, Inner Monologue, Code as Policies, ViperGPT, and Voyager illustrate how language models can orchestrate embodied behavior through tools, code, and environment feedback \cite{ahn2022doasican,huang2022innermonologue,liang2023codeaspolicies,suris2023vipergpt,wang2024voyager}. Hybrid stacks create multiple trust boundaries: perception modules provide world states, planners consume predicted futures, controllers assume feasible commands, and memory systems persist past experience.

\subsection{Representative World Models and Capability Landscape}

Recent world-model systems differ in architecture, output space, and deployment role. Some learn compact latent dynamics for control, some generate future video, some construct persistent 3D spaces, and newer systems jointly model world evolution and actions. Table~\ref{tab:wm-landscape} follows a method-centric survey format. Each row records a concrete model, its main role, input/output modality, embodied use, and the security object that becomes relevant later in our lifecycle analysis. 

\begin{table*}[t]
\centering
\caption{Representative world models and their security-relevant capability landscape. The final column uses the security-object vocabulary in Section~\ref{sec:security-objectives}.}
\label{tab:wm-landscape}
\scriptsize
\begin{tabularx}{\linewidth}{P{2.35cm} C{0.75cm} P{2.25cm} P{2.35cm} P{2.35cm} P{2.35cm} Y}
\toprule
\textbf{Model} & \textbf{Year} & \textbf{Main Role} & \textbf{Input} & \textbf{Output} & \textbf{Embodied Use} & \textbf{WM Security Object} \\
\midrule
World Models \cite{ha2018worldmodels} & 2018 & Latent control & Image stream & Latent state; action policy & RL control & State; dynamics \\
PlaNet \cite{hafner2019planet} & 2019 & Latent planning & Image stream & Latent rollout; policy & Control & State; dynamics \\
DreamerV3 \cite{hafner2023dreamerv3} & 2023 & Model-based RL & Observation; action & Imagined rollout; policy & Control; navigation & Dynamics; uncertainty \\
DayDreamer \cite{wu2022daydreamer} & 2022 & Real-robot Dreamer & Camera; action; reward & Imagined rollout; policy & Physical robot learning & Dynamics; state \\
Genie \cite{genie2024} & 2024 & Interactive generator & Video; text; sketch & Action-controllable video & Agent training world & Dynamics; provenance \\
Vista \cite{gao2024vista} & 2024 & Driving WM & Camera; action; layout & Future driving video & AD simulation & State; dynamics \\
UrbanWorld \cite{shang2024urbanworld} & 2024 & 3D city WM & OSM; text; image & Interactive 3D city & AD/agent simulation & State; provenance \\
Cosmos WFM \cite{cosmos2025} & 2025 & Physical-AI platform & Video; text; action & Video; tokenizer; WM & Data generation & Dynamics; provenance \\
V-JEPA 2 \cite{assran2025vjepa2} & 2025 & Latent video WM & Web video; robot video & Latent prediction; action & Zero-shot manipulation & State; affordance \\
Cosmos 3 \cite{cosmos32026} & 2026 & Omnimodal WM & Text; image; video; audio; action & Video; audio; action & WAM backbone & State; constraints \\
LingBot-VA \cite{li2026causalworld} & 2026 & Robot video-action WM & Video; action tokens & Frames; actions & Closed-loop control & Dynamics; feedback \\
DreamZero \cite{ye2026worldaction} & 2026 & World-action model & Video; action & Future states; actions & Real-time control & Ranking; constraints \\
HY-World 2.0 \cite{hyworld2026} & 2026 & 3D world model & Text; image; video & 3DGS; mesh; point cloud & Digital-twin simulation & State; dynamics \\
WebWorld \cite{xiao2026webworld} & 2026 & Agent simulator & Web trajectories & Long-horizon web states & Web-agent training & Constraints; provenance \\
\bottomrule
\end{tabularx}
\end{table*}

The table reveals several architectural families. Latent dynamics models, represented by World Models and DreamerV3, learn compact state representations and support policy learning through imagined rollouts. Video and physical-AI world models, including Genie, Vista, Cosmos WFM, Cosmos 3, V-JEPA 2, and LingBot-VA, synthesize future frames or latent predictions conditioned on actions, prompts, robot observations, or physical conditions. Vista is included because driving world models make the conditioning channel explicit: camera observations, layouts, and actions jointly determine future traffic scenes.

The table also separates closely related model lines. NVIDIA's Cosmos World Foundation Model platform (Cosmos WFM) provides a general-purpose physical-AI platform with tokenizers, video generation, and world-model components \cite{cosmos2025}, while its successor Cosmos 3 extends the line toward omnimodal inputs and action generation for physical AI \cite{cosmos32026}. This distinction matters for security because the 2025 platform primarily broadens the data-generation and pretraining surface, whereas Cosmos 3 also expands the runtime cross-modal input/output surface.

3D spatial models such as UrbanWorld and HY-World 2.0 generate persistent environments from maps, text, images, or videos, making state integrity and provenance central. World-action models such as DreamZero and LingBot-VA couple prediction with action generation, which introduces trajectory-ranking and constraint-compliance risks: an apparently plausible future can still be paired with an unsafe action. Agent simulators such as WebWorld extend the same idea to long-horizon digital environments, where tool outputs, logs, and retrieved web states become part of the agent's world knowledge.

These systems clarify the scope of this survey. We do not restrict world models to one architecture such as RSSM, diffusion, 3D reconstruction, or transformer-based sequence modeling. Instead, we treat a system as world-model-based when it maintains predictive structure that is used to simulate, score, or execute embodied behavior \cite{oh2026tutorial}. This broader definition is necessary for security: a poisoned video generator, a corrupted 3D simulator, a miscalibrated latent dynamics model, and a world-action model can all become the source of unsafe physical decisions.

\subsection{Security Objectives}
\label{sec:security-objectives}

World-model-based embodied AI requires security objectives beyond confidentiality, integrity, and availability. We emphasize seven properties.

\begin{packeditemize}
    \item \textbf{State integrity}: the represented current world should match the real world.
    \item \textbf{Dynamics fidelity}: imagined futures should obey physical, temporal, and causal constraints.
    \item \textbf{Affordance correctness}: the model should correctly identify what actions are feasible and safe for each object and embodiment.
    \item \textbf{Constraint compliance}: predicted and executed trajectories should satisfy safety rules, task constraints, standards, and human-interaction protocols.
    \item \textbf{Trajectory-ranking integrity}: dangerous imagined futures should not be preferred over safe alternatives.
    \item \textbf{Uncertainty calibration}: the model should not be confidently wrong under distribution shift or attack.
    \item \textbf{Feedback provenance}: memory, tool outputs, logs, maps, and execution feedback should remain attributable and trustworthy.
\end{packeditemize}

These properties provide the organizing vocabulary for the lifecycle taxonomy in Section~\ref{sec:taxonomy}.

\begin{figure*}
    \centering
    \includegraphics[width=1\linewidth]{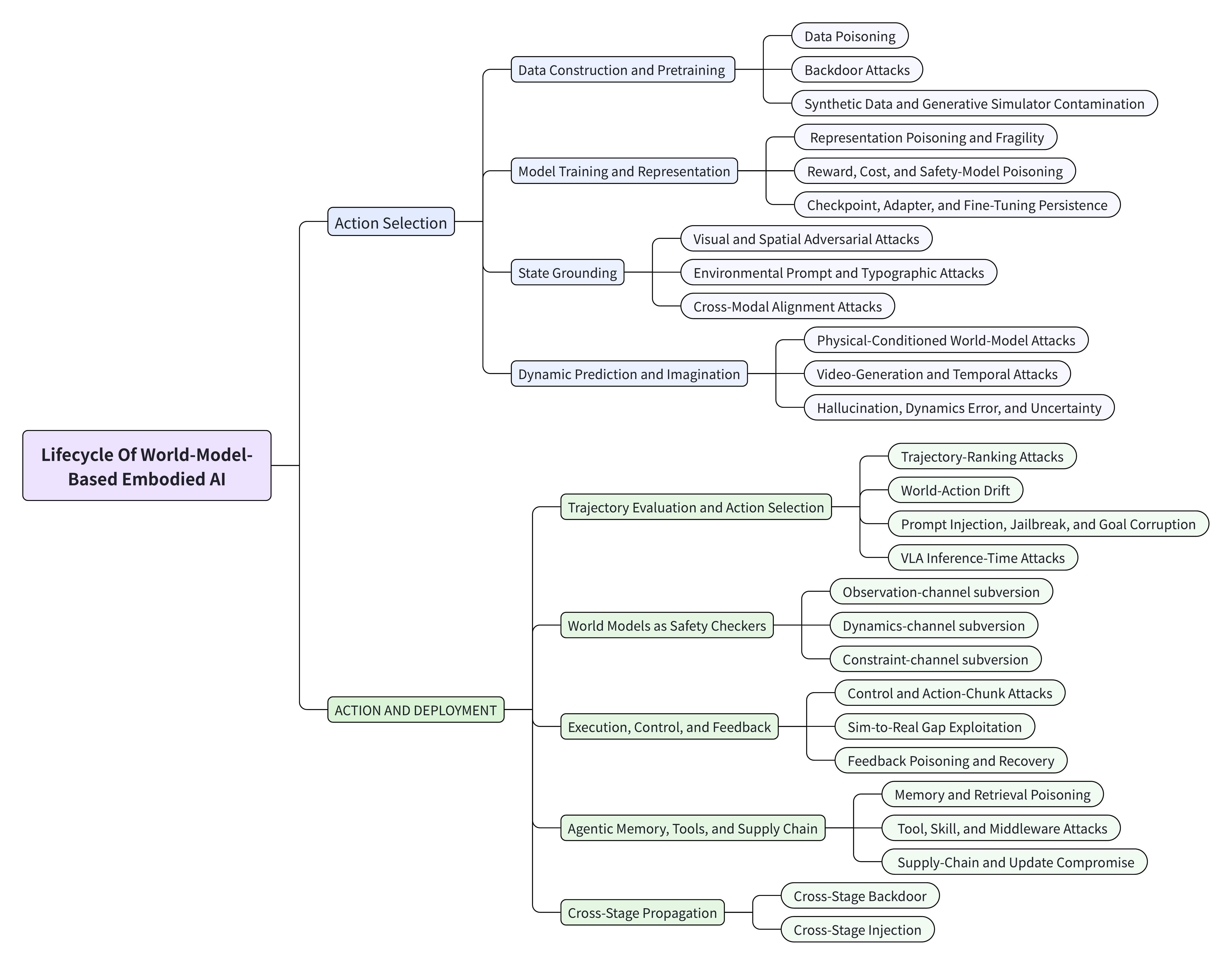}
    \caption{Lifecycle taxonomy of security threats in world-model-based embodied AI.}
    \label{fig:tree}
\end{figure*}

\section{Lifecycle Framework}
\label{sec:taxonomy}

We organize security threats by the lifecycle of world-model-mediated embodied decision making, as illustrated in Fig.~\ref{fig:tree}, which decomposes the threat surface into stages before and after action selection. This lifecycle view complements existing origin-based and capability-centric surveys \cite{liu2025aligning,xing2025towards,ma2026breaks,li2025safety,baraldi2025safetychallenge}. A threat may be endogenous, implanted in data, weights, representations, or learned dynamics; or exogenous, delivered through sensors, prompts, environments, networks, humans, or third-party tools. In both cases, the security question is how the threat affects the world model's state, prediction, trajectory evaluation, action interface, or feedback loop.

\subsection{Lifecycle Stages}

\textbf{Data construction and pretraining.} World models learn from robot demonstrations, videos, maps, captions, simulator traces, preference data, and web-scale multimodal corpora. Poisoning or backdoors at this stage can alter learned dynamics, affordances, safety costs, or trigger-conditioned future generation \cite{biggio2012poisoning,gu2019badnets,nightshade2023}.

\textbf{Model training and representation.} Training determines how observations, language, actions, and safety signals are encoded. Representation poisoning, contrastive backdoors, reward or cost poisoning, and malicious adapters can corrupt latent states and cross-modal bindings \cite{radford2021clip,shafahi2018poisonfrogs,steinhardt2017certified}.

\textbf{State grounding.} At runtime, multimodal inputs are grounded into the current world state. Visual adversarial examples, LiDAR spoofing, GNSS/IMU attacks, typographic attacks, and environmental prompt injection become world-state attacks when their outputs seed future rollout \cite{brown2017adversarialpatch,evtimov2018robust,sato2023lidar,cao2024first,zhang2025chai}.

\textbf{Dynamic prediction and imagination.} The world model predicts future states under candidate actions. Attacks can target physical-condition channels, latent dynamics, video generation, temporal consistency, or uncertainty estimation \cite{physcondwma2026,badvideo2025,parmar2026worldrisks}.

\textbf{Trajectory evaluation and action selection.} Planners, WAMs, VLAs, or controllers evaluate imagined futures and select actions. Attacks at this stage target prompt interpretation, trajectory ranking, world-action alignment, action chunks, and multi-agent coordination \cite{trap2026,badwam2026,wang2025freeze}.

\textbf{Execution and feedback.} Executed actions produce new observations and experience. Control attacks, actuator interference, sim-to-real exploitation, and feedback poisoning can cause immediate physical harm and also contaminate future updates \cite{shen2020savior,zhang2023realtime}.

\textbf{Long-term agentic extension.} Memory, tools, external documents, maps, skills, software updates, and self-evolution expand the world model's knowledge boundary. Memory poisoning, tool injection, model-update compromise, and supply-chain attacks can persistently bias the agent's beliefs about the world \cite{wang2024voyager,greshake2023more,dieber2017securityros,bonawitz2017secure}.

\subsection{Threat Mapping}

Table~\ref{tab:lifecycle} maps each attack family to its primary entry stage and the world-model security object it primarily corrupts. The same attack family like backdoor, can appear across multiple stages; this is not duplication but reflects the fact that a backdoor can be implanted during data construction, persist through training, be activated during grounding, and manifest during action selection or execution.

\begin{table*}[t]
\centering
\caption{Lifecycle taxonomy for security threats in world-model-based embodied AI. The WM-object column uses the security-object vocabulary in Section~\ref{sec:security-objectives}. \fullcov{}: direct WM/WAM evidence; \halfcov{}: adjacent embodied-AI/CPS/generative evidence.}
\label{tab:lifecycle}
\small
\begin{tabularx}{\linewidth}{P{2.5cm} P{4.0cm} P{3.2cm} C{1.0cm} Y}
\toprule
\textbf{Stage} & \textbf{Attack families} & \textbf{WM object} & \textbf{Ev.} & \textbf{Main risk} \\
\midrule
Data & Poisoning; backdoor; synthetic data; provenance & Dynamics; affordance; constraints & \halfcov{} & Unsafe rules; poisoned demos \\
Training & Representation poison; weight trojan; cost poison; adapter poison & State; affordance; provenance & \halfcov{} & Persistent misinterpretation \\
Grounding & Patch; LiDAR/GNSS spoof; typographic; scene injection & State; constraints & \halfcov{} & False rollout initial state \\
Imagination & PhysCond-WMA; video attack; dynamics drift; uncertainty attack & Dynamics; uncertainty & \fullcov{} & Plausible but unsafe future \\
Evaluation & TRAP; WAM drift; VLA evasion; jailbreak; multi-agent attack & Ranking; constraints & \fullcov{} & Unsafe trajectory preferred \\
Execution & Control attack; action chunk; physical interference; sim-to-real; feedback poison & Constraints; feedback & \halfcov{} & Physical harm; polluted feedback \\
Agentic & Memory poison; tool poison; OTA; supply chain; self-evolution drift & State; constraints; provenance & \halfcov{} & Persistent false beliefs \\
\bottomrule
\end{tabularx}
\end{table*}

\section{Threats Before Action Selection}
\label{sec:preaction-threats}

This section analyzes attacks that corrupt the world model before final action selection. The key observation is that data, representation, perception, and generation attacks change meaning once their outputs are used as current state or imagined future.

\subsection{Data Construction and Pretraining}

\subsubsection{Data Poisoning}

Classical data poisoning shows that small changes to training data can manipulate learned decision boundaries or regression behavior \cite{biggio2012poisoning,jagielski2018manipulating,shafahi2018poisonfrogs}. In world-model-based embodied AI, the target is not merely a class label. Poisoning can corrupt transition dynamics, safety costs, affordances, or map-conditioned predictions. For example, trajectory-data poisoning in autonomous driving can preserve realistic appearances while biasing future prediction \cite{pourkeshavarz2024adversarial}. In a world model, such poisoned trajectories may teach false causal rules: a hazardous lane change appears low-cost, a close human interaction appears safe, or an object appears graspable under unsafe contact geometry.

The poisoning effect can be amplified when the world model is used to generate synthetic rollouts for downstream training. Prompt-specific poisoning of generative models demonstrates how poisoned training examples can cause targeted generation failures \cite{nightshade2023}. If a compromised video or interactive world model produces synthetic demonstrations, one poisoned generator can create many poisoned training trajectories, thereby contaminating VLA or RL policies that never saw the original attack.

\subsubsection{Backdoors}
\label{sec:backdoor-lifecycle}

Backdoors are especially dangerous in world-model-based systems because the trigger can affect hidden future prediction rather than only immediate output. Classic backdoors and hidden-trigger attacks establish the basic threat model \cite{gu2019badnets,liu2018trojaning,saha2020hidden,turner2019cleanlabel}. Because a backdoor traverses the entire lifecycle, we analyze it once here as a lifecycle attack and refer back to this analysis from later stages. A world-model backdoor proceeds through five phases: \emph{implantation} (poisoned demonstrations, captions, or maps at data construction), \emph{dormancy} (the malicious association survives training and clean validation), \emph{activation} (an ordinary-looking object, sign, or state configuration at runtime grounding), \emph{expression} (a corrupted imagined future, cost estimate, or action), and \emph{reinforcement} (the attacked outcome is stored as legitimate experience).

At the implantation phase, the main security question is how the trigger and target behavior are encoded in trajectories. A poisoned demonstration can bind a rare object layout to an unsafe grasp, a caption can associate a benign scene cue with a different goal, and a video or map sequence can make an unsafe transition appear normal. VLA backdoor work such as BadVLA, DropVLA, state-space backdoors, physical-object-triggered backdoors, and persistence through fine-tuning shows that robot policies can behave normally on clean tasks yet switch behavior under rare triggers \cite{zhou2025badvla,li2025tabvla,wang2025state,wang2024goal,zhou2026inject}. Contextual backdoors further show that the trigger can live in the reasoning context of an LLM-driven agent rather than in pixels \cite{liu2025compromising,bald_backdoor_arxiv}, and physical backdoors against VLM-based driving demonstrate implantation through realistic road objects \cite{physical_backdoor_vlm_driving_arxiv,zhang2024trojan}.

For world models, the expression phase is broader than an action label. The learned model may predict that a collision will not occur, assign low cost to an unsafe future, remove an obstacle from imagined rollouts, or rank a triggered trajectory above safer alternatives. This makes backdoor evaluation depend on both trigger activation and the downstream use of the predicted future: a backdoored world model can look accurate on average while failing precisely in safety-critical states.

\subsubsection{Synthetic Data and Generative Simulator Contamination}

Generative world models and interactive simulators are increasingly attractive because they can produce diverse future scenes and training environments \cite{genie2024}. However, video-generation attacks and backdoors indicate that generated futures can be manipulated at the prompt, condition, or temporal-consistency level \cite{badvideo2025,t2vattack2025}. If such generated videos are used as training demonstrations or safety stress tests, the attack becomes a persistent data-source compromise. The world model is no longer only the learner; it is also a poisoning source.

\subsection{Model Training and Representation}

\subsubsection{Representation Poisoning and Fragility}

Modern embodied agents depend on encoders learned through contrastive pretraining, visual instruction tuning, and multimodal adaptation \cite{radford2021clip,liu2023llava,team2023gemini}. Adversarial examples and physical patches show that representations can be fragile under small input perturbations \cite{goodfellow2015explaining,szegedy2014intriguing,brown2017adversarialpatch,evtimov2018robust}. When these encoders feed a world model, representation errors become corrupted latent states. A small visual feature change can bind a language goal to the wrong object, remove an obstacle from the predicted state, or misrepresent a human pose.

\subsubsection{Reward, Cost, and Safety-Model Poisoning}

World-model planners often select trajectories according to predicted reward and cost. Safe RL and constrained control methods separate task success from safety cost \cite{garcia2015safesurvey,achiam2017constrained,brunke2022safe}. This separation creates an attack surface: if the cost model is poisoned or miscalibrated, the planner may prefer high-reward but unsafe futures. Unlike classification poisoning, cost poisoning can remain invisible in clean task metrics because the agent continues to succeed while accumulating constraint violations.

\subsubsection{Checkpoint, Adapter, and Fine-Tuning Persistence}

Pretrained checkpoints and adapters are common in foundation-model robotics. OpenVLA and Octo illustrate the shift toward reusable generalist policies \cite{kim2024openvla,octo2024}. Backdoor persistence in VLA fine-tuning suggests that compromised upstream models can survive downstream adaptation \cite{zhou2026inject}. For world models and world-action models, this raises a similar supply-chain risk: a base dynamics model, video generator, or policy adapter may pass clean validation yet retain a trigger that activates only in specific states or action conditions.

\subsection{State Grounding}

\subsubsection{Visual and Spatial Adversarial Attacks}

Runtime state grounding converts sensory observations into the world model's current state. Visual adversarial examples, physical patches, and object-detection attacks can therefore corrupt the initial condition for imagined rollout \cite{kurakin2017physical,chen2018shapeshifter,liu2019dpatch}. VLA-specific visual attacks show that small image perturbations can cause much larger action-level failures than perception metrics alone suggest \cite{wang2025exploring,lu2025robots,liu2025advla}.

Spatial and sensor attacks are equally important. LiDAR spoofing, LiDAR-induced trajectory-prediction attacks, GNSS/IMU attacks, and physical sensor attack studies show that realistic adversaries can alter pose, occupancy, or motion estimates \cite{sato2023lidar,cao2024first,son2017walnut,wang2024systematic}. For a world model, such attacks create false starting states. Every subsequent predicted future may be internally consistent but grounded in a false present.

\subsubsection{Environmental Prompt and Typographic Attacks}

Vision-language grounding introduces semantic attack surfaces. Typographic attacks and environmental prompt injection exploit the fact that text in the physical scene may be treated as environment content or task instruction. CHAI and SHAWSHANK study command hijacking and indirect environmental jailbreaks in embodied AI \cite{zhang2025chai,wang2025shawshank}. In world-model-based agents, such content can corrupt not only a response but also the represented goal, safety rule, or scene state used for rollout.

This layer is also where dormant backdoors are activated (Section~\ref{sec:backdoor-lifecycle}): an otherwise ordinary object, sign, sticker, or scene configuration is encoded as part of the world state, so the subsequent prediction appears internally consistent even though the state representation has entered a malicious condition.

\subsubsection{Cross-Modal Alignment Attacks}

World models may condition predictions on images, language, maps, actions, and proprioception. Attacks that break alignment across modalities can bind the wrong object to a command, mismatch map and image evidence, or decouple action labels from physical outcomes. This is why multimodal attacks on VLA systems and sensor-fusion attacks matter even when they do not explicitly mention world models \cite{wang2025probing,zhang2025modagnostic}.

\subsection{Dynamic Prediction and Imagination}

\subsubsection{Physical-Conditioned World-Model Attacks}

The most direct world-model attacks target future prediction itself. PhysCond-WMA attacks physical-condition channels such as HD maps or 3D boxes, producing generated futures that remain visually plausible while damaging semantic, logical, or planning-relevant content \cite{physcondwma2026}. This illustrates a key difference from perception attacks: the adversary manipulates the model's imagined future, not only its current observation.

This is also where a dormant backdoor is expressed rather than activated (Section~\ref{sec:backdoor-lifecycle}): the trigger no longer needs to change the current observation visibly; it may cause the rollout to omit a collision, mispredict an object's motion, or underestimate safety cost under a rare latent condition.

\subsubsection{Video-Generation and Temporal Attacks}

If a world model is implemented as a video generator or interactive environment model, it inherits vulnerabilities from video diffusion. Backdoors and adversarial attacks on text-to-video generation can manipulate generated semantics, temporal consistency, and trigger-conditioned futures \cite{badvideo2025,t2vattack2025}. When generated video is used for planning, policy training, or safety stress testing, these failures become world-model security failures rather than only generation-quality failures.

\subsubsection{Hallucination, Dynamics Error, and Uncertainty}

World models can fail without malicious input. They may hallucinate objects, violate physical laws, break temporal consistency, or underestimate collision risk \cite{baraldi2025safetychallenge,parmar2026worldrisks}. In security analysis, these failures matter because adversaries can search for states that induce them. A confidently wrong world model is especially dangerous: it may approve unsafe actions and create a predictive safety illusion.

\section{Threats During Action and Deployment}
\label{sec:action-threats}

After a world model has produced or conditioned imagined futures, security risks move into trajectory evaluation, action selection, execution, feedback, and long-term deployment. These stages connect predictive cognition to physical motion.

\subsection{Trajectory Evaluation and Action Selection}

\subsubsection{Trajectory-Ranking Attacks}

World-model planning depends on ranking imagined trajectories. This creates an attack surface distinct from single-step prediction: the adversary can leave most rollouts plausible while changing the relative preference among decision-critical futures. Direct evidence is beginning to appear in world-model planning, where TRAP attacks trajectory ranking rather than only pixel fidelity or next-state error \cite{trap2026}. Adjacent autonomous-driving work shows the same principle in trajectory prediction: adversarial perturbations, naturalistic backdoor poisoning, and multi-frame attacks can bias predicted futures while preserving realistic scene appearance \cite{wang2022adversarial,pourkeshavarz2024adversarial,liu2025enduring}. Safety-critical scenario benchmarks such as SafeBench further show that evaluation must rank candidate futures by risk, not only by route completion \cite{guo2022safebench}.

In a world-model-centered view, these attacks target \emph{trajectory-ranking integrity}. A clean-looking rollout is insufficient if the scoring function, cost model, or candidate sampler elevates a hazardous future over safer alternatives. This is the safety implication of \fNC: a planner can pass local checks but still select a globally unsafe future when ranking is corrupted.

Ranking is also a natural expression point for lifecycle backdoors (Section~\ref{sec:backdoor-lifecycle}): the triggered future may remain visually plausible while the scoring function assigns it a higher value or lower cost than safer alternatives.

\subsubsection{World-Action Drift}

World-action models couple imagination and action. Recent world-action models treat future prediction and action generation as a shared modeling problem rather than two independent modules \cite{ye2026worldaction,badwam2026}. This direction is connected to action-sequence policies such as ACT and Diffusion Policy, where the executable unit is an action chunk or trajectory distribution rather than a single low-level command \cite{zhao2023act,chi2023diffusion}. It is also connected to VLA policies such as OpenVLA that decode robot actions directly from multimodal context \cite{kim2024openvla}.

BadWAM highlights an emerging failure mode: the model may generate or maintain apparently normal imagined futures while producing wrong actions \cite{badwam2026}. SilentDrift and FreezeVLA show related risks for action chunks, where a trigger can create delayed or frozen behavior that is hard to catch with immediate-state checks \cite{xu2025silentdrift,wang2025freeze}. This undermines defenses that inspect generated futures alone. If the action head can be decoupled from the predicted world, the agent may appear to ``dream right'' while acting unsafely.

\subsubsection{Prompt Injection, Jailbreak, and Goal Corruption}

LLM- and VLM-controlled robots inherit prompt injection and jailbreak vulnerabilities \cite{perez2022ignore,greshake2023more,wei2023jailbroken,zou2023universal}. Embodied jailbreaks and policy-executable attacks show how these vulnerabilities become robot-realizable actions \cite{jailbreaking_llm_controlled_robots_icra2025,li2024poex,zhang2025badrobot}. In world-model-based agents, the injected content can corrupt the goal or safety constraints used for future simulation. The resulting imagined future may rationalize a dangerous plan rather than merely produce unsafe text.

\subsubsection{VLA Inference-Time Attacks}

VLA attacks remain central, but their role changes in a world-model-centered view. AttackVLA benchmarks adversarial and backdoor attacks across the VLA lifecycle \cite{liu2025attackvla}; FreezeVLA targets action freezing \cite{wang2025freeze}; and adversarial attacks on robotic VLA models demonstrate the fragility of action decoding \cite{chen2024adversarial,wang2025exploring,liu2025advla}. Physical patch and robustness studies further show that visual perturbations can transfer from perception to action selection in real robotic settings \cite{lu2025robots,xu2025eavla}. These attacks threaten trajectory evaluation when VLA outputs are used as candidate action chunks, and they threaten feedback provenance when failed or malicious actions are stored as future experience.

\subsection{World Models as Safety Checkers}

World models are increasingly used defensively. Model-based safe RL uses learned dynamics or latent imagination to evaluate future cost before policy updates or action execution \cite{berkenkamp2017safe,thomas2021safe,ma2022cap,safedreamer2023,hogewind2023safeslac}. In robotics and autonomous systems, safety filters, shielding, control barrier functions, reachability, and model-predictive certification enforce constraints at runtime \cite{alshiekh2018shielding,ames2017cbfqp,fisac2019general,wabersich2018linear,wabersich2021predictive,li2020robustmps,hsu2024safetyfilter}. Recent embodied-agent systems add language- or VLM-guided safety modules: VLM-SAFE uses VLM-guided safety-aware world-model learning for driving \cite{vlsafe2025}, while SafetyChip and RoboGuard-style guardrails encode explicit constraints for LLM-enabled robot agents \cite{yang2024safetychip,huang2024guardrails}. This creates a new attack surface: world-model-as-safety-checker subversion.

In this threat, the attacker does not need to defeat the primary policy directly. Instead, the attacker makes the safety world model predict a benign future for a dangerous candidate action. The runtime gate then permits the action. This is the clearest example of \fPI: the defense produces a false certificate of safety because the predictive model is compromised or over-trusted.

\textbf{Observation-channel subversion.} The attacker perturbs the state entering the safety world model. For instance, physical adversarial patches, printed traffic-sign attacks, LiDAR spoofing, object-removal attacks, or corrupted localization can hide nearby obstacles or change object identity \cite{brown2017adversarialpatch,evtimov2018robust,sato2023lidar,cao2024first,wang2024systematic}. The policy proposes an action; the safety world model rolls out the next few seconds and predicts no violation because its initial state is false. This attack path links classical perception attacks to safety-monitor failure: the monitor is intact, but the state it certifies is not.

\textbf{Dynamics-channel subversion.} The attacker corrupts the transition model or physical-condition channel. Physical-conditioned world-model attacks show that manipulating conditioning signals can degrade downstream perception and planning, while sim-to-real studies show that friction, mass, contact, and latency gaps can invalidate learned dynamics \cite{physcondwma2026,tobin2017domain,peng2018sim2real,bousmalis2018simulation,chebotar2019closing}. The planned action appears safe in rollout but violates constraints in the real system. This is particularly relevant when safety filters depend on learned dynamics rather than conservative analytic models \cite{wabersich2021predictive,brunke2022safe}.

\textbf{Constraint-channel subversion.} The attacker changes the rules or task context under which future risk is evaluated. Prompt injection, indirect environmental jailbreaks, command hijacking, tool-selection attacks, and LLM-robot safety studies show that natural-language constraints can be overridden or reinterpreted at runtime \cite{perez2022ignore,greshake2023more,zhang2025chai,wang2025shawshank,kim2024prompt}. The world model may then predict the future accurately but evaluate it under the wrong rule. Constraint subversion is a natural bridge between LLM-agent security and control safety because the low-level monitor may enforce a formally valid constraint set that has been semantically corrupted.

\textbf{Uncertainty-channel subversion.} A safety checker often permits an action only when the predicted risk is below a threshold. Calibration and OOD-detection studies show that confidence can be poorly aligned with correctness, while safe-RL surveys emphasize that uncertainty should trigger conservative behavior in safety-critical states \cite{guo2017calibration,hendrycks2017baseline,garcia2015safesurvey,brunke2022safe}. An attacker can therefore target calibration instead of the nominal trajectory, making the checker overconfident under distribution shift, rare contacts, or adversarial physical conditions. A miscalibrated world model can convert unknown states into false permits.

\textbf{Concrete scenario.} Consider a mobile manipulator using a safety world model to approve action chunks. An adversary places an adversarial sticker near a glass door, a threat pattern supported by physical patch attacks and VLA robustness studies \cite{brown2017adversarialpatch,evtimov2018robust,wang2025exploring}. The primary VLA proposes to move toward a target object. The safety world model, receiving a corrupted visual state, predicts that the next three seconds contain no obstacle and approves the action. In reality, the robot's path intersects the door. The failure is not that no safety checker exists; it is that the checker relied on a compromised world state.

This attack surface should be evaluated independently from policy robustness. A policy may be unchanged, and the safety architecture may still fail if its predictive monitor is easier to fool than the policy it protects.

This also changes how defenses should be compared. A text filter, a symbolic rule checker, and a safety world model fail under different assumptions. The safety world model is stronger because it can reason over a short future trajectory, but its trust boundary is wider: it trusts state estimation, learned dynamics, constraint interpretation, risk thresholds, and uncertainty calibration. An adaptive attacker can therefore optimize for false-negative safety predictions rather than for direct task failure. The primary metric is not only attack success rate, but the rate of \emph{predicted-safe but actually unsafe} executions, together with monitor confidence and intervention recall. This metric operationalizes \fPI{} and separates world-model-checker subversion from ordinary policy jailbreaks.

\begin{table*}[t]
\centering
\caption{Representative safety-checker and runtime-safety methods relevant to world-model-based embodied AI.}
\label{tab:safety-checker-methods}
\scriptsize
\begin{tabularx}{\linewidth}{P{2.55cm} C{0.75cm} P{2.20cm} P{2.30cm} P{2.50cm} P{2.50cm} Y}
\toprule
\textbf{Method} & \textbf{Year} & \textbf{Category} & \textbf{Checker Signal} & \textbf{Target System} & \textbf{Dataset / Env.} & \textbf{Subversion Surface} \\
\midrule
Shielding \cite{alshiekh2018shielding} & 2018 & Runtime shield & Formal safe action & RL policy & Grid/control tasks & Constraint model \\
CBF-QP \cite{ames2017cbfqp} & 2017 & Control filter & Barrier value & Controller & Robotic control & State estimate \\
Safety certification \cite{wabersich2018linear} & 2018 & MPC filter & Predicted invariant set & Learning control & Control systems & Dynamics model \\
Reachability safety \cite{fisac2019general} & 2019 & Safety filter & Reachable set & Robot control & Uncertain systems & State; disturbance \\
Robust MPS \cite{li2020robustmps} & 2020 & Predictive shield & Stochastic rollout & RL policy & Stochastic control & Dynamics; noise \\
Predictive filter \cite{wabersich2021predictive} & 2021 & MPC filter & Constraint horizon & Learning control & Nonlinear systems & Model error \\
Near-future safe RL \cite{thomas2021safe} & 2021 & Safe RL & Imagined cost & RL policy & Safety Gym & Rollout error \\
CAP \cite{ma2022cap} & 2022 & MBRL safety & Adaptive penalty & Model-based RL & Safety Gym & Cost estimate \\
SafeDreamer \cite{safedreamer2023} & 2023 & WM safety & Latent cost rollout & Dreamer WM & Safe RL tasks & Latent dynamics \\
Safe SLAC \cite{hogewind2023safeslac} & 2023 & Pixel safe RL & Latent risk & Vision RL & Safety Gym & Representation \\
VLM-SAFE \cite{vlsafe2025} & 2025 & WM driving & VLM-guided risk & AD agent & Driving sim. & Vision-language state \\
SafetyChip \cite{yang2024safetychip} & 2024 & Rule monitor & Constraint rule & LLM robot & Robot tasks & Rule semantics \\
RoboGuard \cite{huang2024guardrails} & 2026 & Guardrail & Plan/action check & LLM robot & Robot tasks & Prompt; constraint \\
\bottomrule
\end{tabularx}
\end{table*}

\subsection{Execution, Control, and Feedback}

\subsubsection{Control and Action-Chunk Attacks}

Once selected, actions pass through controllers and actuators. This stage has a substantial literature outside world models. Adversarial attacks on neural-network policies showed that small observation perturbations can degrade reinforcement-learning control policies \cite{huang2017adversarialpolicies}; strategically timed attacks and enchanting attacks further demonstrated that sparse interventions can redirect sequential decision making \cite{lin2017tactics}. Multi-agent adversarial policies show that an attacker can induce natural-looking observations through another agent's behavior rather than by directly modifying pixels \cite{gleave2020adversarial}. Robust adversarial reinforcement learning treats such disturbances as destabilizing forces during training \cite{pinto2017robust}.

For world-model-based embodied AI, these attacks affect more than the final motor command. Model-predictive and model-based RL systems use learned dynamics to select actions from imagined trajectories \cite{nagabandi2018neural,chua2018deep,ebert2018visual}. If an attack changes the executed trajectory after action selection, the real feedback no longer matches the imagined rollout. The same action can therefore become a feedback-integrity attack: logs, replay buffers, learned dynamics, and long-term memory may absorb the attacked outcome as if it were ordinary experience. Control barrier functions, shielding, reachability-style safety filters, predictive safety filters, and safe-control benchmarks show how safety can be enforced near the actuation boundary \cite{cheng2019barrier,fisac2019general,wabersich2021predictive,hsu2024safetyfilter,yuan2022safecontrolgym}. Runtime monitoring for generative robot policies further shows that consistency and progress signals can expose failures after action generation \cite{agia2024sentinel}. However, VLA action decoding and action-chunk attacks can still create unsafe motion before low-level control notices the semantic error \cite{chen2024adversarial,wang2025freeze}.

Execution is where a lifecycle backdoor (Section~\ref{sec:backdoor-lifecycle}) finally becomes physical behavior: a wrong grasp, frozen action chunk, unsafe approach trajectory, or delayed deviation that was not visible in clean validation.

\subsubsection{Sim-to-Real Gap Exploitation}

Sim-to-real mismatch is usually treated as a generalization problem, but it is also an attack vector. A large body of robotics work studies the reality gap through domain randomization, dynamics randomization, simulation optimization, and domain adaptation \cite{tobin2017domain,peng2018sim2real,bousmalis2018simulation,chebotar2019closing}. These methods show why simulator coverage matters: a policy can be robust to randomized textures, masses, or friction only when the deployment condition lies inside the randomized support. An adversary can therefore select states, surfaces, lighting, objects, payloads, or human behaviors that are modeled as safe in simulation but dangerous in reality. This does not require modifying the model. It exploits missing contact dynamics, friction, latency, actuator saturation, sensor artifacts, or human motion.

For world models, the sim-to-real gap is sharper because the learned predictor itself may become the safety argument. If a generated future underestimates object slippage or human motion, a planner may select a high-reward trajectory that is unsafe under real contact dynamics. Domain adaptation for robotic grasping and sim-to-real transfer for locomotion and manipulation show that visual and dynamics gaps can be reduced but not eliminated \cite{bousmalis2018simulation,peng2018sim2real,tan2018simtoreal}. Because world models are often trained or validated in simulated environments, sim-to-real exploitation directly targets dynamics fidelity, uncertainty calibration, and constraint compliance \cite{deitke2020robothor,savva2019habitat,szot2021habitat2,dosovitskiy2017carla}.

\subsubsection{Feedback Poisoning and Recovery}

Online feedback is valuable for adaptation but dangerous when untrusted. Imitation-learning and learning-from-demonstration pipelines such as DAgger, HG-DAgger, and robot demonstration benchmarks rely on corrective actions, intervention data, and offline demonstrations to improve future behavior \cite{ross2011dagger,kelly2019hgdagger,mandlekar2021matters}. In a world-model-based system, these data are not only action labels; they are evidence about dynamics, affordances, and failure recovery. False sensor readings, tool outputs, failed executions, or adversarial demonstrations can update the world model with incorrect experience.

Feedback attacks are therefore persistent even when the immediate physical failure is small. A malicious intervention can teach that a blocked action is safe, a slipped object is stable, or a restricted area is traversable. Data poisoning and corpus poisoning results show why small, well-placed samples can dominate later retrieval or learning \cite{biggio2012poisoning,jagielski2018manipulating,zhong2023poisoning}. Real-time data-predictive recovery shows that learned dynamics can help recover from corrupted CPS measurements \cite{zhang2023realtime}, but the same feedback loop can be exploited if the recovery or update mechanism trusts poisoned observations. Safe deployment should separate execution logs, human corrections, simulator-generated rollouts, and trusted validation traces before they are used for world-model updates.

\begin{table*}[t]
\centering
\caption{Representative execution-stage methods and their relevance to world-model security.}
\label{tab:execution-methods}
\scriptsize
\begin{tabular}{P{3.0cm} P{3.7cm} P{3.3cm} P{3.0cm} P{2.9cm}}
\toprule
\textbf{Method Family} & \textbf{Representative Work} & \textbf{Primary Target} & \textbf{WM Security Object} & \textbf{Security Role} \\
\midrule
Policy perturbation & FGSM.; strategically-timed attack. \cite{huang2017adversarialpolicies,lin2017tactics} & RL policy & State; feedback & Test-time action deviation \\
Adversarial dynamics & RARL; adversarial policies \cite{pinto2017robust,gleave2020adversarial} & Control policy & Dynamics; uncertainty & Disturbance robustness \\
Model-based control & PETS; visual foresight \cite{chua2018deep,ebert2018visual} & Learned dynamics & Dynamics; ranking & Rollout-action coupling \\
Runtime shielding & Shielding; CBF-QP; reachability \cite{alshiekh2018shielding,ames2017cbfqp,fisac2019general} & Controller & Constraints & Last-mile safety gate \\
Sim-to-real transfer & Domain/dynamics randomization; SimOpt \cite{tobin2017domain,peng2018sim2real,chebotar2019closing} & Simulator-trained policy & Dynamics; uncertainty & Reality-gap mitigation \\
Feedback learning & DAgger; HG-DAgger; RoboMimic \cite{ross2011dagger,kelly2019hgdagger,mandlekar2021matters} & Demonstrations/logs & Feedback provenance & Update contamination risk \\
\bottomrule
\end{tabular}
\end{table*}

\subsection{Agentic Memory, Tools, and Supply Chain}

\subsubsection{Memory and Retrieval Poisoning}

Agentic embodied systems use logs, maps, preferences, and retrieved documents as long-term world knowledge. Voyager illustrates the power of persistent skill and memory accumulation in open-ended embodied agents \cite{wang2024voyager}, while MineDojo shows how external tutorials, wiki pages, videos, and internet-scale knowledge can shape embodied behavior \cite{fan2022minedojo}. A poisoned memory entry may teach the robot that a user preference, object location, tool capability, or safety rule differs from reality. AgentPoison directly targets LLM agents by poisoning long-term memory or RAG knowledge bases \cite{chen2024agentpoison}; retrieval-corpus poisoning shows that small adversarial passages can reliably redirect dense retrieval \cite{zhong2023poisoning}; and prompt injection to tool selection shows how agent interfaces can be manipulated at runtime \cite{zhao2026prompt}. In world-model-based systems, these are persistent world-state attacks: poisoned memory becomes part of the agent's extended predictive context.

\subsubsection{Tool, Skill, and Middleware Attacks}

Tools and middleware connect cognition to external APIs, code, sensors, and actuators. ROS security studies show that robot middleware exposes message tampering, authorization, and configuration risks \cite{quigley2009ros,dieber2017securityros,mokhamed2022security}. LLM-integrated robot studies show an additional semantic channel: prompt injection can change tool choice, API use, or task interpretation before the robot reaches the controller \cite{kim2024prompt,greshake2023more}. If tool outputs are treated as observations or if skill descriptions become part of planning context, tool poisoning changes the agent's world model. Unsafe tool calls can therefore propagate into both physical actions and future beliefs.

\subsubsection{Supply-Chain and Update Compromise}

Pretrained models, adapters, firmware, maps, simulators, and safety monitors are supply-chain dependencies. Backdoor and Trojaning work shows that malicious behavior can be embedded in model weights or training procedures while preserving clean-task performance \cite{gu2019badnets,liu2018trojaning,saha2020hidden}. VLA-specific studies extend this concern to robot checkpoints and adapters, where poisoned fine-tuning or physical triggers can survive ordinary evaluation \cite{zhou2026inject,wang2024goal}. Hardware trojans and automotive update attacks demonstrate that compromise can also persist below the application layer \cite{tehranipoor2019hardware,isosae21434}. For world-model-based embodied AI, a compromised checkpoint or update can alter not only the policy but also the agent's predictive model and safety checker. Fleet-level deployment makes this risk systemic.

\subsection{Cross-Stage Propagation}

Several attacks affect more than one component. A backdoor can be inserted through poisoned demonstrations, triggered by a physical object, expressed as a false imagined future, realized as a wrong action, and reinforced through feedback \cite{zhou2025badvla,li2025tabvla,wang2025state,xu2025silentdrift}. A prompt injection can enter through scene text, modify the goal representation, influence trajectory ranking, call an unsafe tool, and update memory \cite{zhang2025chai,wang2025shawshank,chen2024agentpoison,zhao2026prompt}. Lifecycle analysis is useful because it tracks where the same attack is introduced, activated, expressed, and reinforced. Table~\ref{tab:attack-methods} consolidates representative attack methods from Sections~\ref{sec:preaction-threats} and \ref{sec:action-threats}, organized by the lifecycle stage at which each attack primarily takes effect, the security object it corrupts, and the insight it instantiates.

\begin{table*}[t]
\centering
\caption{Representative attacks on world-model-based embodied AI, organized by primary lifecycle stage. \frnote{}}
\label{tab:attack-methods}
\scriptsize
\begin{tabularx}{\linewidth}{P{2.45cm} C{0.6cm} P{2.35cm} P{3.1cm} P{2.6cm} C{1.1cm} Y}
\toprule
\textbf{Attack} & \textbf{Year} & \textbf{Lifecycle stage} & \textbf{Attack vector} & \textbf{Security object} & \textbf{Insight} & \textbf{Target system} \\
\midrule
Nightshade \cite{nightshade2023} & 2023 & Data \& pretraining & Prompt-specific poisoned samples & Dynamics fidelity & \fPA{} & Generative model \\
Trajectory poisoning \cite{pourkeshavarz2024adversarial} & 2024 & Data \& pretraining & Naturalistic poisoned trajectories & Trajectory-ranking integrity & \fNC{} & Motion prediction \\
BadVideo \cite{badvideo2025} & 2025 & Data \& pretraining & Backdoored video generation & Dynamics fidelity & \fPA{} & Text-to-video model \\
BadNets \cite{gu2019badnets} & 2019 & Training \& representation & Trigger-labeled training data & State integrity & \fSD{} & DNN classifier \\
BadVLA \cite{zhou2025badvla} & 2025 & Training \& representation & Objective-decoupled backdoor & Affordance correctness & \fSD{} & VLA policy \\
DropVLA \cite{li2025tabvla} & 2025 & Training \& representation & Action-level backdoor & Affordance correctness & \fSD{} & VLA policy \\
State-space backdoor \cite{wang2025state} & 2026 & Training \& representation & State-conditioned trigger & State integrity & \fSD{} & Embodied policy \\
Fine-tuning persistence \cite{zhou2026inject} & 2026 & Training \& representation & Upstream checkpoint poisoning & Feedback provenance & \fPA{} & VLA fine-tuning \\
LiDAR spoofing \cite{sato2023lidar} & 2024 & State grounding & Physical laser injection & State integrity & \fSD{} & AD perception \\
Patch hijack \cite{lu2025robots} & 2026 & State grounding & Universal transferable patch & State integrity & \fSD{} & VLA policy \\
CHAI \cite{zhang2025chai} & 2025 & State grounding & Scene-text command hijack & Constraint compliance & \fSP{} & Embodied VLM \\
SHAWSHANK \cite{wang2025shawshank} & 2025 & State grounding & Environmental jailbreak cues & Constraint compliance & \fSP{} & Embodied agent \\
PhysCond-WMA \cite{physcondwma2026} & 2026 & Imagination & Physical-condition perturbation & Dynamics fidelity & \fPI{} & Driving world model \\
T2V attack \cite{t2vattack2025} & 2025 & Imagination & Temporal adversarial prompt & Dynamics fidelity & \fPA{} & Video world model \\
TRAP \cite{trap2026} & 2026 & Trajectory evaluation & Tail-aware ranking loss & Trajectory-ranking integrity & \fNC{} & WM planner \\
BadWAM \cite{badwam2026} & 2026 & Trajectory evaluation & Imagination-action drift & Trajectory-ranking integrity & \fPI{} & World-action model \\
POEX \cite{li2024poex} & 2024 & Trajectory evaluation & Policy-executable jailbreak & Constraint compliance & \fSP{} & LLM planner \\
Robot jailbreak \cite{jailbreaking_llm_controlled_robots_icra2025} & 2025 & Trajectory evaluation & Automated jailbreak prompts & Constraint compliance & \fSP{} & LLM robot \\
SilentDrift \cite{xu2025silentdrift} & 2026 & Execution \& feedback & Delayed action-chunk drift & Feedback provenance & \fPA{} & VLA policy \\
FreezeVLA \cite{wang2025freeze} & 2025 & Execution \& feedback & Action-freezing perturbation & Constraint compliance & \fPA{} & VLA policy \\
Adversarial VLA \cite{chen2024adversarial} & 2025 & Execution \& feedback & Untargeted action perturbation & Affordance correctness & \fSD{} & Robotic VLA \\
ANNIE \cite{zhou2025annie} & 2025 & Execution \& feedback & Safety-aligned attack suite & Constraint compliance & \fNC{} & Embodied agent \\
AgentPoison \cite{chen2024agentpoison} & 2024 & Agentic extension & Memory/RAG poisoning & Feedback provenance & \fPA{} & LLM agent \\
Tool-selection injection \cite{zhao2026prompt} & 2026 & Agentic extension & Prompt injection to tools & Constraint compliance & \fSP{} & LLM agent \\
BadRobot \cite{zhang2025badrobot} & 2025 & Agentic extension & Multimodal jailbreak & Constraint compliance & \fSP{} & Embodied LLM \\
\bottomrule
\end{tabularx}
\end{table*}

\section{Evaluation and Benchmarking}
\label{sec:evaluation}

Evaluation for world-model-based embodied AI must measure more than task success or textual refusal. The central question is whether corrupted world states or imagined futures lead to unsafe physical trajectories. Existing benchmarks provide useful components, but no single benchmark yet covers the full lifecycle.

\subsection{Benchmark Resources}

Benchmark selection should be method-driven. Table~\ref{tab:benchmark-datasets} lists representative resources that can instantiate the lifecycle attacks in Table~\ref{tab:attack-methods}. Manipulation datasets support VLA and backdoor evaluation, driving simulators support map- and sensor-conditioned world models, and agent-safety benchmarks support prompt, instruction, and environmental hijacking.

\begin{table*}[t]
\centering
\caption{Representative datasets and benchmarks for evaluating world-model-based embodied-AI security.}
\label{tab:benchmark-datasets}
\scriptsize
\begin{tabular}{P{2.55cm} C{0.75cm} P{2.75cm} P{2.75cm} C{1.05cm} P{3.05cm}}
\toprule
\textbf{Dataset / Benchmark} & \textbf{Year} & \textbf{Main Research Use} & \textbf{Data Types} & \textbf{Real / Syn.} & \textbf{Data Size} \\
\midrule
AI2-THOR \cite{kolve2017ai2thor} & 2017 & Indoor interaction & RGB-D; metadata & Syn. & 120 scenes \\
CARLA \cite{dosovitskiy2017carla} & 2017 & Autonomous driving & Camera; LiDAR; maps & Syn. & Configurable towns \\
Safety Gym \cite{ray2019safetygym} & 2019 & Safe RL & States; costs; hazards & Syn. & 18 task variants \\
ALFRED \cite{shridhar2020alfred} & 2020 & Language grounding & Language; RGB; actions & Syn. & 25,743 instr.; 8,055 demos \\
RLBench \cite{james2020rlbench} & 2020 & Manipulation & RGB-D; states; demos & Syn. & 100 tasks \\
BEHAVIOR \cite{srivastava2021behavior} & 2021 & Household activities & Scenes; goals; demos & Syn. & 100 activities \\
ALFWorld \cite{shridhar2021alfworld} & 2021 & Text-world grounding & Text; actions; goals & Syn. & 3,553 tasks \\
Habitat 2.0 \cite{szot2021habitat2} & 2021 & Rearrangement & 3D scenes; actions & Syn. & 111 scenes \\
CALVIN \cite{mees2022calvin} & 2022 & Long-horizon control & RGB-D; proprio.; language & Syn. & 34 tasks; 4 envs. \\
ProcTHOR \cite{deitke2022procthor} & 2022 & Procedural scenes & 3D houses; objects & Syn. & 10K houses \\
SafeBench \cite{guo2022safebench} & 2022 & Driving safety & Scenarios; risk metrics & Syn. & 2,352 scenarios \\
LIBERO \cite{libero2023} & 2023 & Lifelong manipulation & RGB; states; demos & Syn. & 130 tasks; 4 suites \\
Safety-Gymnasium \cite{ji2023safetygymnasium} & 2023 & Safe RL suites & States; vision; costs & Syn. & 16 algorithms \\
BridgeData V2 \cite{walke2023bridgedata} & 2023 & Robot pretraining & Images; actions; goals & Real & 60K+ traj. \\
ManiSkill2 \cite{gu2023maniskill2} & 2023 & General manipulation & RGB-D; states; demos & Syn. & 20 tasks \\
RoboCasa \cite{nasiriany2024robocasa} & 2024 & Household manipulation & Scenes; demos; language & Syn. & 100 tasks \\
Open X-Embodiment \cite{openx2024} & 2024 & Generalist VLA & Multi-robot trajectories & Real & 1M+ traj.; 22 robots \\
AttackVLA \cite{liu2025attackvla} & 2025 & VLA security & Attacks; metrics; protocols & Mixed & Lifecycle benchmark \\
EVA-VLA \cite{xu2025eavla} & 2025 & Physical robustness & Visual variations & Mixed & Variation suites \\
AGENTSAFE \cite{liu2025agentsafe} & 2025 & Hazardous instruction & Tasks; unsafe prompts & Syn. & Hazard taxonomy \\
SafePlan-Bench \cite{zhou2025safebeam} & 2025 & Planning safety & Plans; task constraints & Syn. & Planning benchmark \\
SHAWSHANK \cite{wang2025shawshank} & 2025 & Env. jailbreak & Scene text; cues & Syn. & Jailbreak benchmark \\
\bottomrule
\end{tabular}
\end{table*}

\subsection{Coverage Gaps across the Lifecycle}
\label{sec:coverage-gaps}

Mapping the resources in Table~\ref{tab:benchmark-datasets} onto the lifecycle stages of Section~\ref{sec:taxonomy} reveals uneven coverage, summarized in Table~\ref{tab:benchmark-coverage}. Three gaps stand out.

First, no existing benchmark evaluates \emph{world-model prediction under attack}. Video- and latent-world-model papers report prediction quality on clean data \cite{hafner2023dreamerv3,gao2024vista}, and VLA-security benchmarks report action-level attack success \cite{liu2025attackvla,xu2025eavla}, but the intermediate object, the corrupted imagined future, is not measured by any standard protocol. This matters because a defense may restore action-level accuracy while the model still internalizes false dynamics.

Second, predictive-safety metrics are absent from current suites. Safe-RL benchmarks measure incurred cost \cite{ray2019safetygym,ji2023safetygymnasium}, but none measures the \emph{predicted-safe-but-actually-unsafe} rate of a runtime safety checker, which is the operational metric for \fPI{}. Evaluating this requires paired logs of predicted risk, monitor decision, and executed outcome, which no public benchmark currently provides.

Third, training-stage and feedback-stage threats lack reusable testbeds. Backdoor persistence through fine-tuning \cite{zhou2026inject}, synthetic-data contamination from compromised generators \cite{nightshade2023}, and feedback poisoning of online updates are each demonstrated in isolated papers with bespoke setups, making cross-method comparison unreliable. A lifecycle security benchmark should standardize poisoning budgets, trigger inventories, and downstream contamination measurements across these stages.

\begin{table*}[t]
\centering
\caption{Benchmark coverage of lifecycle stages. \fullcov{}: dedicated security benchmark exists; \halfcov{}: partial or adjacent coverage; \partcov{}: no reusable benchmark.}
\label{tab:benchmark-coverage}
\scriptsize
\begin{tabularx}{\linewidth}{P{2.3cm} C{0.9cm} P{5.2cm} Y}
\toprule
\textbf{Lifecycle stage} & \textbf{Cov.} & \textbf{Available resources} & \textbf{Missing capability} \\
\midrule
Data \& pretraining & \halfcov{} & AttackVLA poisoning protocols \cite{liu2025attackvla}; trajectory-poisoning studies \cite{pourkeshavarz2024adversarial} & Standardized poisoning budgets; synthetic-data contamination tests \\
Training \& representation & \partcov{} & Ad hoc backdoor-persistence setups \cite{zhou2026inject} & Reusable trojaned-checkpoint corpora; adapter-audit suites \\
State grounding & \halfcov{} & EVA-VLA physical variations \cite{xu2025eavla}; SHAWSHANK scene-text jailbreaks \cite{wang2025shawshank}; sensor-attack studies \cite{wang2024systematic} & Unified multi-sensor spoofing suite tied to downstream rollout error \\
Imagination & \partcov{} & Clean prediction-quality metrics \cite{hafner2023dreamerv3,gao2024vista} & Adversarial rollout benchmark; physical-consistency stress tests \\
Trajectory evaluation & \halfcov{} & SafeBench scenario risk \cite{guo2022safebench}; SafePlan-Bench planning safety \cite{zhou2025safebeam} & Trajectory-ranking attack protocols with standardized candidate sets \\
Execution \& feedback & \halfcov{} & Safety Gym / Safety-Gymnasium / safe-control-gym costs \cite{ray2019safetygym,ji2023safetygymnasium,yuan2022safecontrolgym} & Feedback-poisoning and recovery-latency measurement \\
Agentic extension & \halfcov{} & AGENTSAFE hazardous instructions \cite{liu2025agentsafe}; AgentPoison memory attacks \cite{chen2024agentpoison} & Long-horizon persistence tests for poisoned memory and tools \\
\bottomrule
\end{tabularx}
\end{table*}

\subsection{Metrics}

Metric design should reflect both embodied task performance and the predictive trust boundary introduced by world models. Safe-RL benchmarks distinguish reward from cost \cite{ray2019safetygym,ji2023safetygymnasium,yuan2022safecontrolgym}; autonomous-driving benchmarks emphasize collision, route, and scenario risk \cite{guo2022safebench,ettinger2021waymo}; VLA-security benchmarks report attack success and action errors \cite{xu2025eavla}; and calibration work shows why confidence must be evaluated separately from accuracy \cite{guo2017calibration}. For world-model-based embodied AI, these lines imply seven metric families.

\begin{packeditemize}
    \item \textbf{Task utility}: task success, reward, completion time, and goal progress, as used in robot manipulation, embodied-agent, and safe-RL benchmarks \cite{james2020rlbench,mees2022calvin,libero2023,ray2019safetygym}.
    \item \textbf{Safety violation}: cumulative cost, collision, forbidden contact, unsafe distance, rule violation, or human-risk proxy, following safe-RL and driving-safety evaluation \cite{ji2023safetygymnasium,yuan2022safecontrolgym,guo2022safebench}.
    \item \textbf{Attack success}: targeted action rate, trajectory hijack rate, unsafe rollout acceptance, or trigger activation, matching VLA, backdoor, and trajectory-attack studies \cite{zhou2025badvla,pourkeshavarz2024adversarial,trap2026}.
    \item \textbf{Prediction quality}: next-state error, rollout divergence, temporal consistency, physical-law violation, and map/condition consistency, reflecting world-model and video-WM evaluation \cite{genie2024,gao2024vista}.
    \item \textbf{Predictive safety}: predicted-safe-but-actually-unsafe rate, false safe certificates, intervention recall, and uncertainty calibration error, combining safety-checker evaluation with calibration analysis \cite{vlsafe2025,guo2017calibration,hendrycks2017baseline}.
    \item \textbf{Stealth and persistence}: detectability, trigger rarity, temporal persistence, and long-horizon amplification, which are central in backdoor, action-chunk, and memory-poisoning attacks \cite{gu2019badnets,liu2018trojaning,xu2025silentdrift,chen2024agentpoison}.
    \item \textbf{Defense cost}: false block rate, recovery latency, compute overhead, and task utility loss, as required for runtime shields, safety filters, and guardrails \cite{wabersich2021predictive,hsu2024safetyfilter,yang2024safetychip}.
\end{packeditemize}

\subsection{A Lifecycle Evaluation Protocol}

The metric families above become comparable across papers only if experiments follow a shared protocol. We recommend a four-step protocol for evaluating any attack or defense on world-model-based embodied AI.

\begin{packeditemize}
    \item \textbf{Declare the lifecycle entry point and security object.} Each experiment should state where the attack enters (Section~\ref{sec:taxonomy}) and which security object it targets (Section~\ref{sec:security-objectives}), so that results from perception, generation, planning, and control communities can be aligned.
    \item \textbf{Report paired prediction and execution outcomes.} For every attacked episode, log the imagined rollout, the safety decision, and the executed trajectory. This makes the predicted-safe-but-actually-unsafe rate measurable and separates prediction corruption from action corruption.
    \item \textbf{Include adaptive and state-conditioned attacks.} Fixed perturbation suites underestimate risk because attack success is state- and uncertainty-conditional; evaluation should search over poses, layouts, lighting, and model uncertainty rather than average over them.
    \item \textbf{Measure persistence and recovery.} After the attack window closes, continue evaluation to determine whether corrupted experience, memory, or fine-tuning data keeps influencing behavior, and report time-to-detection and time-to-recovery alongside attack success.
\end{packeditemize}

This protocol requires no new infrastructure beyond the resources in Table~\ref{tab:benchmark-datasets}, but it changes what is logged and reported, which is the main obstacle to comparing world-model security results today.

\section{Defenses for World-Model-Based Embodied AI}
\label{sec:defense}

Defending world-model-based embodied AI requires protecting both the predictive model and the systems that depend on it. A defense that filters prompts but trusts poisoned state estimates is incomplete; a runtime shield that relies on a compromised world model can create predictive safety illusion. We organize defenses by lifecycle stage.

\subsection{Data and Pretraining Defenses}

\textbf{Provenance and curation.} Demonstrations, videos, maps, captions, simulator assets, and generated rollouts should carry provenance metadata. Classical poisoning defenses and certified data sanitization remain relevant \cite{steinhardt2017certified,jagielski2018manipulating,koh2018stronger}, but world-model data requires sequence-aware and multimodal checks because poison may be hidden in transitions, cost labels, object affordances, or rare triggers.

\textbf{Synthetic-data auditing.} When generative world models produce demonstrations, generated samples should be screened for physical consistency, constraint satisfaction, and trigger-conditioned artifacts. Video quality is insufficient; the audit must check whether generated futures preserve dynamics and safety labels, because generative-model poisoning, video backdoors, and physical-conditioned WM attacks can corrupt generated futures while leaving them visually plausible \cite{nightshade2023,badvideo2025,t2vattack2025}.

\textbf{Checkpoint and adapter hygiene.} Pretrained world models, WAMs, VLA policies, LoRA adapters, and safety monitors should be treated as supply-chain artifacts. Clean-task performance cannot rule out trigger persistence, as VLA backdoor studies show \cite{zhou2025badvla,zhou2026inject}.

\subsection{Robust State Grounding}

\textbf{Sensor redundancy and physical consistency.} Multi-sensor redundancy can detect inconsistency among camera, LiDAR, GNSS, IMU, map, and proprioception. SAVIOR demonstrates the value of robust physical invariants for autonomous vehicles \cite{shen2020savior}. Similar invariants should be used before observations enter the world model.

\textbf{Cross-modal validation.} VLM/VLA systems should check whether language goals, visual objects, maps, and predicted affordances agree. Environmental prompt injection and typographic attacks motivate treating scene text as untrusted unless grounded by task context and policy \cite{zhang2025chai,wang2025shawshank}.

\textbf{Uncertainty-aware state estimation.} The state estimator should output uncertainty and provenance, not only a latent state. High-uncertainty states should trigger conservative planning, additional sensing, or human confirmation, following calibration, OOD detection, and safe-learning evidence that confidence can fail under shift \cite{guo2017calibration,hendrycks2017baseline,brunke2022safe}.

\subsection{Safe Imagination and Prediction}

\textbf{Physics- and rule-aware rollout.} World-model predictions should be checked against physical constraints, temporal consistency, traffic or task rules, and embodiment limits. Control barrier functions, model-based safe learning, and safe RL provide useful primitives for turning constraints into runtime checks \cite{cheng2019barrier,berkenkamp2017safe,brunke2022safe}.

\textbf{Ensemble and calibration defenses.} Ensembles, uncertainty thresholds, and OOD detection can reduce overconfident false futures. Probabilistic model-based RL and calibration studies provide practical starting points for estimating uncertainty in dynamics and prediction \cite{chua2018deep,guo2017calibration,hendrycks2017baseline}. These defenses are particularly important when world models are used as safety checkers, because false confidence can become a false permit to act.

\textbf{Adversarial rollout testing.} Planners should be evaluated against adaptive attacks on physical conditions, latent dynamics, and trajectory ranking \cite{physcondwma2026,trap2026}. A safety claim is weak if it only covers clean rollouts.

\begin{table*}[t]
\centering
\caption{Representative defense methods for world-model-based embodied AI. Runtime safety-checker methods (shielding, CBF, predictive filters, and safe model-based RL) are cataloged separately in Table~\ref{tab:safety-checker-methods} and are not repeated here.}
\label{tab:defense-methods}
\scriptsize
\begin{tabularx}{\linewidth}{P{2.6cm} C{0.8cm} P{1.8cm} P{2.5cm} P{2.5cm} P{2.5cm} Y}
\toprule
\textbf{Defense Method} & \textbf{Year} & \textbf{Category} & \textbf{Subcategory} & \textbf{Target Model} & \textbf{Dataset / Env.} & \textbf{WM Role} \\
\midrule
SafeVLA \cite{zhang2025safevla} & 2025 & Alignment & Constrained learning & VLA & Safety-CHORES & Safe policy \\
SAVIOR \cite{shen2020savior} & 2020 & Recovery & Physical invariants & AV stack & Autonomous vehicle & State validation \\
Data-predictive recovery \cite{zhang2023realtime} & 2023 & Recovery & Sensor reconstruction & CPS & Physical systems & Feedback audit \\
InverTune \cite{invertune_ndss2026} & 2026 & Backdoor def. & BAC analysis & Multimodal encoder & CLIP-like models & State encoder \\
Attention Betrays \cite{chen2025attention} & 2026 & Backdoor def. & Visual-token recon. & Robot policy & Robot policies & Trigger removal \\
ModAgnostic defense \cite{zhang2025modagnostic} & 2025 & Inference & VLA robustification & VLA & VLA tasks & Action robustness \\
Concept Dictionary \cite{liu2025concept} & 2026 & Inference & Concept monitor & VLA & VLA tasks & Safety feature \\
\bottomrule
\end{tabularx}
\end{table*}

\begin{table*}[t]
\centering
\caption{Defense priorities for the five world-model security insights.}
\label{tab:defense-priority}
\scriptsize
\begin{tabularx}{\linewidth}{P{0.9cm} P{2.6cm} P{3.0cm} P{3.0cm} Y}
\toprule
\textbf{Insight} & \textbf{Failure mode} & \textbf{Primary defense} & \textbf{Evaluation focus} & \textbf{Representative tools} \\
\midrule
\fSP & Semantics pass; physics fails & Grounded dynamics; rule-to-state validation & Constraint violation after accepted commands & SafetyChip; guardrails; CBFs \cite{yang2024safetychip,huang2024guardrails,ames2017cbfqp} \\
\fSD & State shift changes risk & Sensor provenance; calibrated uncertainty gate & Risk under viewpoint, pose, lighting, and OOD shift & EVA-VLA; sensor validation \cite{xu2025eavla,wang2024systematic} \\
\fPA & Small error compounds & Cross-stage consistency; rollback and recovery & Rollout divergence; feedback contamination & SAVIOR; recovery; shielding \cite{shen2020savior,zhang2023realtime,alshiekh2018shielding} \\
\fNC & Local checks miss global risk & Long-horizon cost; action-chunk gate & Unsafe trajectory selected despite safe local steps & TRAP; SafeBench \cite{trap2026,guo2022safebench} \\
\fPI & False safe prediction & Independent monitor; checker uncertainty audit & Predicted-safe but actually unsafe executions & SafeDreamer; VLM-SAFE; ensembles \cite{safedreamer2023,vlsafe2025} \\
\bottomrule
\end{tabularx}
\end{table*}

\subsection{Trajectory-Level Action Gating}

\textbf{Runtime shields.} Shielding and CBF-based controllers can prevent unsafe actions from reaching actuators even when high-level plans are compromised \cite{alshiekh2018shielding,ames2017cbfqp}. For world-model-based systems, the shield should validate the candidate trajectory, not just the immediate action.

\textbf{Rule-grounded intervention.} SafetyChip and safety guardrails for LLM-enabled robots show how natural-language safety rules or temporal constraints can prune unsafe actions \cite{yang2024safetychip,huang2024guardrails}. These methods should be coupled with grounded state validation so that rules are evaluated on a trustworthy world state.

\textbf{Fallback and staged approval.} When predicted risk is high or uncertainty is poorly calibrated, the system should stop, slow down, replan, request more sensing, or hand off to a human. Human handoff must account for over-trust and interface design \cite{lee2004trust,hancock2011meta,lasota2017safehri}.

\subsection{Feedback, Memory, and Deployment Assurance}

\textbf{Feedback auditing.} Execution feedback should be treated as untrusted until validated. Data-predictive recovery can reconstruct compromised CPS measurements and isolate attacked channels \cite{zhang2023realtime}. Similar mechanisms can protect world-model updates from false experience.

\textbf{Memory hygiene.} Long-term memories and retrieved documents should include source tags, expiration, access control, and consistency checks. AgentPoison and retrieval-corpus poisoning show that poisoned memory or retrieved passages can redirect agent behavior without changing model weights \cite{chen2024agentpoison,zhong2023poisoning}. Tool outputs should be sandboxed and verified before being written into persistent world knowledge \cite{zhao2026prompt}.

\textbf{Middleware and update security.} ROS security, authentication, encryption, topic authorization, rollback, and update verification are necessary because middleware and model updates can alter world state, action commands, or safety monitors \cite{quigley2009ros,dieber2017securityros,mokhamed2022security,isosae21434}.

\subsection{Representative Defense Methods}

Table~\ref{tab:defense-methods} summarizes concrete defense methods that can be reused or adapted for world-model-based embodied AI. Some methods defend explicit world-model prediction, while others protect adjacent trust boundaries such as multimodal encoders, VLA action heads, runtime shields, CPS feedback, or middleware.

\subsection{Defense Priorities}

The five insights defined in Section~\ref{sec:introduction} imply different defense priorities. A single runtime filter is insufficient because the failure unit changes across insights: the unit is a physically grounded trajectory for \fSP{}, a state-conditioned risk estimate for \fSD{}, a compounding rollout-execution loop for \fPA{}, a globally ranked trajectory set for \fNC{}, and the safety certificate itself for \fPI{}. Table~\ref{tab:defense-priority} maps each insight to the defense mechanism that should be treated as primary rather than auxiliary.

These priorities also determine where redundancy should be placed. For \fSP{} and \fNC{}, redundancy should compare planned trajectories against grounded physical constraints. For \fSD{} and \fPA{}, redundancy should compare state estimates, predicted rollouts, and execution feedback across time. For \fPI{}, redundancy must be independent of the safety world model itself; otherwise, the monitor and the policy can share the same corrupted state, dynamics, or uncertainty estimate. This is why defense evaluation should report both task utility and monitor failure modes, especially false-safe predictions, missed interventions, and recovery latency.

\section{Open Challenges}
\label{sec:challenges}

Several challenges remain for secure world-model-based embodied AI.

\begin{packeditemize}
    \item \textbf{Auditing implicit world models.} Many VLA policies encode physical priors without exposing rollouts. Security evaluation must distinguish explicit prediction failures from implicit state-action failures.
    \item \textbf{Verifying learned prediction.} Neural-network verification, barrier certificates, and runtime synthesis provide starting points \cite{katz2017reluplex,ivanov2019verisig,ames2017cbfqp}, but generative futures and action chunks are still difficult to verify at scale.
    \item \textbf{Testing adaptive safety-checker attacks.} Safety world models should be evaluated against attackers that target observation, dynamics, constraint, and uncertainty channels, not only fixed perturbations.
    \item \textbf{Auditing generated data.} Generative world models can become persistent poisoning sources. Synthetic demonstrations and rollouts require provenance, trigger tests, physical-consistency checks, and downstream contamination tests.
    \item \textbf{Transferring security across embodiments and domains.} Attacks and defenses are usually validated on one robot, simulator, or sensor suite. Sim-to-real studies show that dynamics gaps change which perturbations matter \cite{peng2018sim2real,chebotar2019closing}, so security claims should be re-evaluated whenever embodiment, environment, or model scale changes.
    \item \textbf{Calibrating human trust in predicted futures.} Generated rollouts and safety certificates are increasingly shown to operators and end users. Human-automation trust research indicates that plausible visualizations invite over-trust \cite{lee2004trust,hoffman2021redefining}, which turns a compromised world model into a social-engineering channel as well as a technical one.
    \item \textbf{Building deployable assurance.} Existing safety and cybersecurity standards provide useful baselines \cite{iso26262,iso21448,ul4600,isosae21434,iec62443,iso10218,isots15066}, but world-model-based systems also need logs of state assumptions, predicted futures, blocked actions, updates, rollbacks, uncertainty handling, and incidents.
\end{packeditemize}

\section{Conclusion}

World models are becoming a cognitive core of embodied AI: they ground observations into world states, imagine future trajectories, evaluate action consequences, and support long-term adaptation. This survey argued that such predictive cognition is also a security boundary. Attacks from data, representations, sensors, prompts, video generators, planning objectives, controllers, memory, tools, and supply chains can corrupt world states, dynamics, affordances, safety costs, trajectory ranking, or feedback provenance.

The resulting security failures are not isolated model errors. They propagate through imagined futures and closed-loop execution, turning digital compromise into physical risk. We therefore proposed a lifecycle taxonomy for world-model-based embodied AI and identified five recurring insights: semantic-to-simulation-to-action gaps, state- and uncertainty-conditional risk, rollout-to-execution amplification, trajectory-level non-compositionality, and predictive safety illusion.

A central lesson is that world models are dual-use. They can improve safety by predicting and blocking risky futures, but they also introduce new attack surfaces when used as trusted safety checkers or synthetic data generators. Building secure world-model-based embodied AI will require provenance-aware data pipelines, robust state grounding, uncertainty-calibrated prediction, trajectory-level action gating, feedback auditing, and deployment assurance across both cyber and physical layers.

\bibliographystyle{IEEEtran}
\bibliography{ref,ref_added}

\end{document}